\documentclass[
 reprint,
 amsmath,amssymb,
 aps,
 prd,       
 floatfix,  
]{revtex4-2}

\usepackage[T1]{fontenc}
\usepackage{graphicx}
\usepackage{bm}
\usepackage{color}
\usepackage{booktabs}
\usepackage{placeins}

\begin{document}

\title{Optical Appearances and Multiple Lensed Rings of Self-Interacting Boson Stars with a Double-Well Potential}

\author{Zefang Fan$^{1}$}
\author{Sen Guo$^{2}$}
\author{Xianggao Wang$^{1,*}$}
\affiliation{
$^{1}$School of Physical Science and Technology, Guangxi University, Nanning 530004, People's Republic of China\\
$^{2}$College of Physics and Electronic Engineering, Chongqing Normal University, Chongqing 401331, People's Republic of China
}
\email{Z. Fan: 2307301022@st.gxu.edu.cn; S. Guo: sguophys@126.com; X. Wang: wangxg@gxu.edu.cn}

\date{\today}

\begin{abstract}
As horizon-scale imaging opens a new era of strong-field gravity tests, self-interacting boson stars have emerged as compelling dark matter candidates and natural black hole mimickers. We systematically investigate the optical appearances and multiple lensed rings of such horizonless objects illuminated by a geometrically thin accretion disk. To overcome the severe numerical stiffness caused by the slowly decaying scalar tail, we introduce a robust hybrid metric reconstruction framework---seamlessly bridging an interior 6-parameter Pad\'e approximant to the exact exterior Schwarzschild spacetime via a globally $C^2$-continuous cubic spline. This ensures high-precision ray-tracing across the entire domain. For the adopted double-well potential, we identify a strict upper bound ($\phi_0 \lesssim 0.317$) for the existence of regular static solutions. Crucially, we uncover a striking non-monotonic evolution in the shadow size, driven by a macroscopic mass-amplitude turnover that is fundamentally rooted in the microphysical competition between attractive and repulsive self-interactions. Within the most compact configurations, extreme gravitational lensing generates prominent nested higher-order photon rings and distinctly separated secondary images. Furthermore, exploring the scalar potential parameter space ($\Lambda, \Gamma$) reveals a distinct diagonal gradient in the shadow size. Our findings demonstrate that next-generation VLBI observations can move beyond merely identifying horizonless objects, offering a novel geometric avenue to directly probe the microphysical couplings of dark matter scalar fields.
\end{abstract}

\maketitle

\section{Introduction}
\label{sec:intro}

\par
The Event Horizon Telescope (EHT) collaboration has delivered horizon-scale images of the supermassive black holes in M87* and Sagittarius A*, opening a new era of strong-field gravity tests \cite{EHT2019,EHT2022}. This achievement represents a triumph of Very Long Baseline Interferometry (VLBI), which has now emerged as a precision instrument for probing extreme gravitational environments \cite{EHT2019,EHT2022,Zensus2026,EHT2019VI}. While these observations have traditionally been interpreted within the framework of general relativity, they also motivate a crucial theoretical question: what alternative compact objects could produce shadow-like images? In particular, scalar-field solitons have been widely discussed as horizonless dark-matter candidates that can still possess photon spheres and therefore cast shadows \cite{Cardoso2019,Bertone2005,Vagnozzi2024}. Among the most compelling such candidates are boson stars---solitonic solutions of a complex scalar field minimally coupled to gravity. Being horizonless and singularity-free, they nevertheless act as natural ``black-hole mimickers'' \cite{Kaup1968,Ruffini1969,Liebling2017,Cunha2017}.

\par
Over the past years, the landscape of boson-star imaging has been dramatically enriched by various accretion scenarios. Thin-disk studies have revealed that boson stars are chameleonic; they can closely mimic black-hole shadows while simultaneously exposing their unique identities through distinguishable inner light-ring structures, polarization signatures, and multiple bright rings \cite{Li2026,Wang2022a,Rosa2023,Zeng2025Pol}. Parallel investigations have examined the shadows of related compact objects with scalar fields and noncommutative geometries, further demonstrating how additional microphysical hair alters the shadow geometry \cite{Cunha2015,Zeng2022a}. Polarimetric ring images of black holes, such as those from the EHT, provide an essential observational benchmark for comparing these exotic objects \cite{Narayan2021,Akiyama2021Pol}. This exploration has also extended beyond standard general relativity: boson-star solutions have been constructed in modified gravity frameworks such as Palatini $f(R)$ gravity \cite{MasoFerrando2021}, while imaging studies have been extended to exotic couplings such as nonlinear electrodynamics \cite{Zeng2025,Ye2026} and alternative bosonic fields such as Proca stars \cite{Brito2016,Herdeiro2021}. Meanwhile, shadow templates from modified gravity have become essential tools for constraining gravity \cite{Chen2021a,Chen2021b,Kuang2021,Kuang2022,ChenBin2020,ChenBin2021}, and lensing studies have shown that the light rings of rotating boson stars can differ significantly from black-hole shadows \cite{Huang2025ParityOdd}. Furthermore, the motion of test particles in rotating boson star backgrounds has provided crucial insights into their surrounding accretion dynamics \cite{Zhang2022}. Yet, beneath this rich tapestry of macroscopic images, a fundamental driving force is frequently overlooked: the macroscopic properties of boson stars are not arbitrary, but are fundamentally sculpted by their microphysical scalar-field potential \cite{Wang2022a,He2025}.

\par
To prevent the scalar field from dispersing and to balance gravitational collapse, self-interactions are essential. A standard model in effective field theories, such as for Q-balls and self-interacting dark matter, requires a potential with both attractive and repulsive terms \cite{Coleman1985,Friedberg1976,Volkov2002,Marsh2016,Boskovic2022,Gomez2024,deSa2025}. The attractive quartic term binds the field into a soliton, while the repulsive sextic term stabilizes the star against collapse, naturally leading to a double-well potential. The central scientific problem we address here is: How does the microscopic competition between these attractive and repulsive interactions imprint itself on the macroscopic shadow geometry?

\par
Answering this question fundamentally requires accurately resolving the background metric. However, this poses severe numerical challenges. The repulsive sextic interaction produces a slowly decaying scalar tail, meaning the spacetime metric approaches the Schwarzschild form only gradually at large radii. Consequently, any naive matching between the interior and exterior metrics introduces derivative discontinuities. These discontinuities force adaptive geodesic integrators to take vanishingly small steps, frequently causing convergence failures \cite{Grould2016}. While previous ray-tracing studies either integrated the metric numerically over the entire domain \cite{Wang2022a,He2025} or employed a single global fitting formula, both approaches become prohibitively expensive or inaccurate when dealing with the slow decay of a self-interacting field.

\par
In this paper, we overcome these numerical bottlenecks by developing a hybrid metric reconstruction framework. We fit the interior spacetime with a 6-parameter Pad\'e approximant to capture the rapid curvature variations inside the star. This interior solution is then smoothly matched to the exact exterior Schwarzschild metric, as dictated by Birkhoff's theorem \cite{Birkhoff1923}, through a globally $C^2$-continuous cubic spline interpolation. Employing this robust ray-tracing method alongside a geometrically thin, optically thin accretion disk \cite{Luminet1979,Shakura1973}, we systematically investigate the optical appearances of self-interacting boson stars. The disk emission is characterized by a Gaussian-like profile, and the observed specific intensity consistently incorporates both gravitational redshift and Keplerian Doppler boosting \cite{Page1974,Cunningham1973}.

\par
Our analysis uncovers a set of highly non-trivial observational signatures that challenge conventional intuition. While one might naively expect the shadow to grow monotonically with the central field amplitude, we demonstrate a striking non-monotonic behavior driven by the mass-amplitude turnover arising from complex microphysical interactions. This breaks the conventional paradigm and opens a new theoretical window to probe the nature of dark matter. Within the most compact regime of this parameter space, nested higher-order photon rings become distinctly visible inside the shadow \cite{Gralla2020,Johnson2020}, furnishing a geometric probe of compactness. Furthermore, under varying observation inclinations, the shadow reveals intricate multiple lensed rings, featuring both nested photon rings and distinctly separated secondary images. Recent theoretical studies have highlighted that such multi-ring structures serve as precise fingerprints for distinguishing black holes from alternative compact objects \cite{Guo2024KN,Gong2026Shadows,Gralla2019}. Perhaps most compellingly, varying the scalar potential parameters generates a distinct diagonal gradient in the shadow size across the parameter space. These distinctive morphological fingerprints suggest a radical observational prospect: by decoding the fine structure of horizonless compact objects with next-generation VLBI, we can directly probe the microphysics of dark matter scalar fields---a feat once thought impossible from black hole shadows alone.

\par
The remainder of this paper is organized as follows. In Sec.~\ref{sec:framework}, we present the theoretical framework and the hybrid metric reconstruction method. In Sec.~\ref{sec:method}, we describe the ray-tracing algorithm and the thin-disk accretion model. In Sec.~\ref{sec:results}, we present the numerical results for the shadow images. Finally, in Sec.~\ref{sec:conclusions}, we summarize our findings and discuss the observational implications.

\section{Theoretical Framework}
\label{sec:framework}

\subsection{Einstein-Klein-Gordon Equations and Numerical Solution}
\label{sec:ekg}

\par
We consider a system in which a complex scalar field is minimally coupled to gravity. The dynamics of the system are governed by the action (in natural units $G=c=1$):
\begin{equation}
    \mathcal{S} = \int d^4x \sqrt{-g} \left[ \frac{R}{2\kappa} - g^{\mu\nu} \partial_\mu \Phi^* \partial_\nu \Phi - V(|\Phi|^2) \right],
\end{equation}
where $\kappa = 8\pi$, $R$ is the Ricci scalar, and $\Phi$ is the complex scalar field with a general potential \(V(|\Phi|^2)\). Varying the action with respect to the metric \(g_{\mu\nu}\) and the scalar field \(\Phi^*\) yields the coupled Einstein field equations and the Klein-Gordon (KG) equation:
\begin{equation}
    R_{\mu\nu} - \frac{1}{2} g_{\mu\nu} R = \kappa T_{\mu\nu}, \quad \nabla_\mu \nabla^\mu \Phi = \frac{\partial V}{\partial \Phi^*},
\end{equation}
where the energy-momentum tensor of the scalar field is given by
\begin{equation}
    T_{\mu\nu} = \partial_\mu \Phi^* \partial_\nu \Phi + \partial_\nu \Phi^* \partial_\mu \Phi - g_{\mu\nu} \left( g^{\alpha\beta} \partial_\alpha \Phi^* \partial_\beta \Phi + V(|\Phi|^2) \right).
\end{equation}

\par
To describe a static, spherically symmetric boson star, we adopt the standard metric ansatz
\begin{equation}
    ds^2 = -A(r) dt^2 + B(r)^{-1} dr^2 + r^2 (d\theta^2 + \sin^2\theta d\varphi^2),
    \label{eq:metric_ansatz}
\end{equation}
where the metric functions \(A(r)\) and \(B(r)\) depend only on the radial coordinate \(r\). The scalar field is assumed to be a monochromatic oscillating solution,
\begin{equation}
    \Phi(r,t) = \phi(r) e^{-i\omega t},
\end{equation}
with \(\phi(r)\) representing the real radial profile and \(\omega\) being the eigenfrequency of the field. Substituting this ansatz into the energy-momentum tensor, we obtain its non-zero mixed components:
\begin{align}
    T^t_t &= -B \phi'^2 - \frac{\omega^2 \phi^2}{A} - V, \\
    T^r_r &= B \phi'^2 - \frac{\omega^2 \phi^2}{A} - V, \\
    T^\theta_\theta &= T^\varphi_\varphi = -B \phi'^2 + \frac{\omega^2 \phi^2}{A} - V,
\end{align}
where a prime denotes the derivative with respect to \(r\).

\par
By projecting the Einstein field equations along the $tt$ and $rr$ components, i.e., $G^t_t = \kappa T^t_t$ and $G^r_r = \kappa T^r_r$, we can extract the differential equations for the metric functions $A(r)$ and $B(r)$. Similarly, the KG equation governs the evolution of the scalar field. To facilitate numerical integration, we introduce the auxiliary variable \(p(r) = d\phi/dr\), which reduces the second-order Klein-Gordon equation to a first-order system. This recasts the coupled equations into a set of four first-order ordinary differential equations (ODEs):
\begin{align}
    \frac{dA}{dr} &= \frac{A}{B r} \left( 1 - B + r^2 \kappa V \right) + \frac{r \kappa \phi^2 \omega^2}{B} + A r \kappa p^2, \\
    \frac{dB}{dr} &= \frac{1-B}{r} - \frac{r \kappa}{A} \left( A V + \phi^2 \omega^2 \right) - B r \kappa p^2, \\
    \frac{d\phi}{dr} &= p, \\
    \frac{dp}{dr} &= -\frac{1}{2} \left( \frac{1}{A}\frac{dA}{dr} + \frac{1}{B}\frac{dB}{dr} + \frac{4}{r} \right) p - \frac{1}{B} \left( \frac{\omega^2}{A} - \frac{1}{2\phi} \frac{\partial V}{\partial \phi} \right) \phi.
\end{align}

\par
To solve this ODE system, appropriate boundary conditions must be imposed. At the center of the star ($r \to 0$), regularity of the geometry and the scalar field requires \(A(0)=1\) (as a gauge choice for the initial guess), \(B(0)=1\), \(\phi(0)=\phi_0\), and \(p(0)=0\). The physical central metric is determined after an overall rescaling (as described below). At spatial infinity ($r \to \infty$), asymptotic flatness requires the scalar field to vanish and the metric to approach the Minkowski form, i.e., $\phi(\infty)=0$ and $A(\infty)=B(\infty)=1$. For numerical stability, the integration is initialized at a small cutoff $r = 10^{-5}$ with the corresponding central values to avoid the coordinate singularity at the origin.

\par
In general relativity, the total mass of an asymptotically flat spacetime is defined by the Arnowitt-Deser-Misner (ADM) mass \cite{Arnowitt1962}. For our static, spherically symmetric spacetime, the total ADM mass $M$ of the boson star can be extracted from the asymptotic behavior of the metric function $B(r)$ at large radii:
\begin{equation}
    M = \lim_{r \to \infty} \frac{r}{2} (1 - B(r)).
\end{equation}
We employ the standard shooting method to search for the eigenfrequency \(\omega\) such that the global solution simultaneously satisfies both the central regularity and the asymptotic flatness conditions. Numerically, the ADM mass is evaluated at the cutoff radius \(r_1\) as \(M \simeq \frac{r_1}{2}(1 - B(r_1))\), and this same value is consistently used for the exterior Schwarzschild matching. After obtaining the numerical solution, an overall rescaling of \(A(r)\) and \(\omega\) is performed to ensure that \(g_{tt} \to -1\) at spatial infinity. The physical central metric component after rescaling is denoted as \(A_0 \equiv -g_{tt}(0)\).

\subsection{The Scalar Field Potential and Critical Points}
\label{sec:potential}

\par
To construct regular non-topological solitons (boson stars), the scalar field potential must incorporate both attractive and repulsive self-interactions. As shown in the classic works on Q-balls and solitonic stars \cite{Volkov2002, Collodel2022}, the minimal renormalizable polynomial potential that allows for stable bound states without gravity is a sextic potential. We adopt a double-well potential:
\begin{equation}
    V(\phi) = \mu^2 \phi^2 - \Lambda \phi^4 + \Gamma \phi^6,
\end{equation}
where $\mu$ represents the scalar mass, and $\Lambda$ and $\Gamma$ correspond to the attractive and repulsive self-interaction constants, respectively. The quartic term ($-\Lambda \phi^4$) provides the attractive force that binds the scalar field into a compact object, while the repulsive sextic term ($\Gamma \phi^6$) provides the outward pressure necessary to prevent the soliton from collapsing into a singularity.

\par
This potential possesses a rich structure of extrema. By setting the first derivative of the potential to zero,
\begin{equation}
    \frac{\partial V}{\partial \phi} = 2\mu^2\phi - 4\Lambda\phi^3 + 6\Gamma\phi^5 = 2\phi(\mu^2 - 2\Lambda\phi^2 + 3\Gamma\phi^4) = 0,
\end{equation}
we find that the term in the parentheses can be viewed as a quadratic equation in $x = \phi^2$. For the potential to exhibit a ``double-well'' shape (i.e., an absolute global minimum and a central local barrier), the discriminant of this quadratic equation must be positive, yielding the condition $\Lambda^2 > 3\Gamma\mu^2$. To clearly exhibit the double-well characteristics while ensuring a regular ground state, we fix the benchmark parameters to $\mu = 1$, $\Lambda = 50$, and $\Gamma = 625$, which safely satisfy this condition. Solving the equation yields three critical physical extrema: a local minimum at $\phi_0 = 0$ (the symmetric vacuum), a local maximum at $\phi_0 = \sqrt{1/75} \approx 0.1155$ (the potential barrier), and a global minimum at $\phi_0 = 0.20$ (the absolute vacuum expectation value, VEV).

\par
\begin{figure}[htbp]
\centering
\includegraphics[width=\columnwidth]{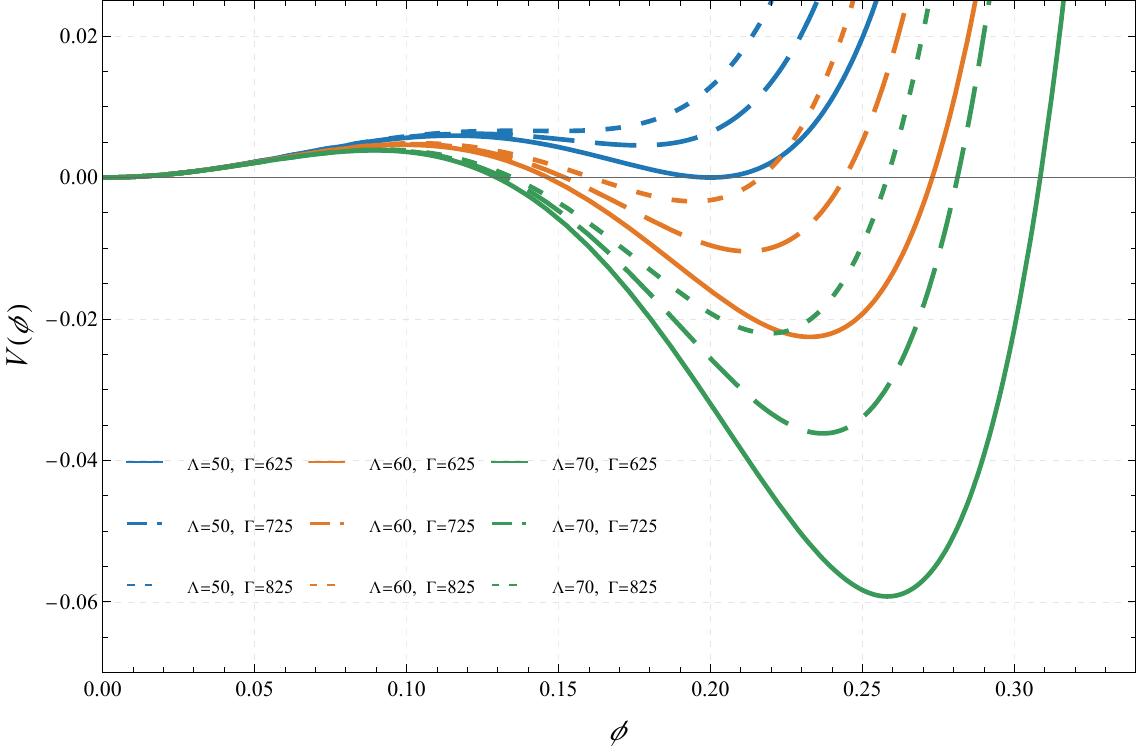}
\caption{The shape of the double-well potential $V(\phi)$ for different combinations of self-interaction parameters. The color scheme (blue, orange, green) distinguishes the attractive parameter $\Lambda \in \{50, 60, 70\}$, while the line style (solid, dashed, dotted) indicates the repulsive parameter $\Gamma \in \{625, 725, 825\}$.}
\label{fig:potential_curves}
\end{figure}

To visually illustrate how the self-interaction parameters modify the potential landscape, we plot the potential curves for various combinations of $\Lambda$ and $\Gamma$ in Fig.~\ref{fig:potential_curves}. As will be demonstrated in subsequent sections, the variations in these potential parameters directly dictate the macroscopic shadow images of the boson stars.

\par
The existence of regular boson stars is highly sensitive to the position of the central amplitude \(\phi_0\) relative to these extrema. For the benchmark potential parameters, the value \(\phi_0=0.20\) corresponds to the absolute vacuum expectation value (VEV) of the potential, representing a typical ground-state configuration. Therefore, in our subsequent inclination scan, we set the central amplitude to this VEV (\(\phi_0 = 0.20\)) to represent the gravitational lensing signatures of a physically stable ground state. In the later parameter sensitivity analysis (Sec.~\ref{sec:sensitivity}), \(\phi_0\) will be dynamically adjusted to the corresponding VEV for each parameter pair.

\subsection{Hybrid Metric Reconstruction and Smooth Interpolation Algorithm}
\label{sec:fitting}

\par
By varying the central amplitude $\phi_0$ and fixing the potential parameters, a discrete set of boson star solutions can be numerically obtained. While semi-analytical approaches have been proposed to construct boson star wavefunctions in specific limits \cite{Kling2017}, direct numerical integration to spatial infinity for relativistic self-interacting stars remains computationally prohibitive due to the slowly decaying exponential tail of the scalar field. We therefore terminate the numerical integration at a cutoff radius \(r_1\), where the scalar field has essentially decayed and the metric derivative is sufficiently small (typically $|A'(r_1)| \sim 10^2/M$).

\par
To perform high-precision backward ray-tracing, the metric must remain highly smooth and physical across the entire domain, including the observer's location at $r=200M$. We propose a hybrid metric reconstruction and smooth interpolation scheme. First, for the interior region $r \in [0.01M, 0.9r_1]$, we sample 500 points and fit the numerical metric components $g_{tt} = -A(r)$ and $g_{rr} = B(r)^{-1}$ using a least-squares method with a 6-parameter rational Pad\'e approximant. Unlike standard Taylor polynomials, the Pad\'e approximant is much more effective at capturing the rapid variations and curvature of the metric inside the star. The specific fitting functions are given by
\begin{align}
    -g_{tt}(r) = A(r) &\approx \frac{c_{0tt} + c_{1tt} r + c_{2tt} r^2 + c_{3tt} r^3}{1 + c_{4tt} r + c_{5tt} r^2}, \label{eq:pade_gtt} \\
    g_{rr}(r) = B(r)^{-1} &\approx \frac{c_{0rr} + c_{1rr} r + c_{2rr} r^2 + c_{3rr} r^3}{1 + c_{4rr} r + c_{5rr} r^2}, \label{eq:pade_grr}
\end{align}
where the choice of a cubic numerator and quadratic denominator ensures sufficient degrees of freedom to achieve a fit quality of $R^2 > 0.999$ for both metric components. The fitted coefficients ($c_{ntt}$ and $c_{nrr}$) for various central amplitudes are detailed in Tables~\ref{tab:pade_gtt} and \ref{tab:pade_grr}.

\begin{table*}[htbp]
\centering
\caption{Numerical values of the central amplitude $\phi_0$, ADM mass $M$, eigenfrequency $\omega$, and the corresponding 6-parameter Pad\'e fitting coefficients for the metric function $g_{tt}$. The fit quality is quantified by $R^2_{tt}$. Note that $c_{0tt}$ can become slightly negative for the most compact configurations due to the extremely small central value of $A(r)$; the global $R^2$ remains excellent.}
\label{tab:pade_gtt}
\renewcommand{\arraystretch}{1.3}
\small
\begin{tabular}{l c c c c c c c c c}
\toprule
$\phi_0$ & $M$ & $\omega$ & $R^2_{tt}$ & $c_{0tt}$ & $c_{1tt}$ & $c_{2tt}$ & $c_{3tt}$ & $c_{4tt}$ & $c_{5tt}$ \\
\midrule
0.06 & 0.50833 & 0.859838 & 0.99996 & 0.5171 & 0.01408 & 0.1034 & 0.0000464 & 0.06553 & 0.1062 \\
0.09 & 0.48372 & 0.811133 & 0.99993 & 0.3778 & -0.003352 & 0.2106 & 0.0000778 & 0.1228 & 0.2153 \\
0.12 & 0.45682 & 0.767532 & 0.99992 & 0.2624 & -0.04621 & 0.3398 & 0.000169 & 0.1499 & 0.3482 \\
0.15 & 0.42687 & 0.732707 & 0.99989 & 0.1681 & -0.1092 & 0.5219 & 0.0002632 & 0.1881 & 0.5341 \\
0.19 & 0.368 & 0.726977 & 0.99989 & 0.0799 & -0.2115 & 0.9393 & 0.0004608 & 0.2895 & 0.9578 \\
0.22 & 0.31966 & 0.76836 & 0.99993 & 0.04146 & -0.2747 & 1.565 & 0.0003282 & 0.5628 & 1.58 \\
0.25 & 0.29928 & 0.812276 & 0.99999 & 0.01793 & -0.2214 & 2.149 & 0.0001758 & 0.9751 & 2.157 \\
0.28 & 0.30711 & 0.82542 & 0.99999 & 0.002723 & -0.01093 & 2.012 & \(2.525 \times 10^{-6}\) & 1.197 & 2.013 \\
0.31 & 0.31942 & 0.818402 & 0.99999 & -0.002609 & 0.1617 & 1.452 & 0.00004517 & 1.036 & 1.455 \\
\bottomrule
\end{tabular}
\end{table*}

\begin{table*}[htbp]
\centering
\caption{Numerical values of the central amplitude $\phi_0$, ADM mass $M$, eigenfrequency $\omega$, and the corresponding 6-parameter Pad\'e fitting coefficients for the metric function $g_{rr}$. The fit quality is quantified by $R^2_{rr}$. The Pad\'e fit remains excellent even when $A(0)$ is extremely small.}
\label{tab:pade_grr}
\renewcommand{\arraystretch}{1.3}
\small
\begin{tabular}{l c c c c c c c c c}
\toprule
$\phi_0$ & $M$ & $\omega$ & $R^2_{rr}$ & $c_{0rr}$ & $c_{1rr}$ & $c_{2rr}$ & $c_{3rr}$ & $c_{4rr}$ & $c_{5rr}$ \\
\midrule
0.06 & 0.50833 & 0.859838 & 0.99910 & 0.9845 & -0.309 & 0.09294 & -0.00001204 & -0.3722 & 0.09175 \\
0.09 & 0.48372 & 0.811133 & 0.99956 & 0.9787 & -0.4062 & 0.1718 & -0.00004534 & -0.5201 & 0.1691 \\
0.12 & 0.45682 & 0.767532 & 0.99956 & 0.973 & -0.4937 & 0.2784 & -0.0000913 & -0.6747 & 0.2737 \\
0.15 & 0.42687 & 0.732707 & 0.99965 & 0.9678 & -0.5538 & 0.4382 & -0.0001294 & -0.8419 & 0.432 \\
0.19 & 0.368 & 0.726977 & 0.99970 & 0.9602 & -0.4259 & 0.9348 & 0.00003028 & -1.083 & 0.9342 \\
0.22 & 0.31966 & 0.76836 & 0.99931 & 0.951 & 0.4087 & 2.073 & 0.000511 & -1.109 & 2.093 \\
0.25 & 0.29928 & 0.812276 & 0.99798 & 0.9322 & 3.985 & 4.999 & 0.002556 & -0.003948 & 5.096 \\
0.28 & 0.30711 & 0.82542 & 0.99682 & 0.9553 & 16.67 & 11.63 & 0.00652 & 5.711 & 11.95 \\
0.31 & 0.31942 & 0.818402 & 0.99623 & 0.3672 & 385.9 & 157.9 & 0.164 & 186.3 & 166.2 \\
\bottomrule
\end{tabular}
\end{table*}

\par
We note that the Pad\'e approximant is strictly used only on this fitting interval. Near the origin, the physical metric is retained from the exact numerical solution. For highly compact configurations (e.g., $\phi_0 = 0.31$), the coefficient $c_{0tt}$ can become slightly negative. This is a well-known artefact of rational fitting and is not used as the physical $A(0)$; the ray-tracing procedure relies on the globally splined metric rather than the bare Pad\'e formula.

\par
For the exterior region $r > r_1$, the scalar field has decayed to negligible values, and the spacetime is well approximated by a vacuum. According to Birkhoff's theorem \cite{Birkhoff1923}, we match the interior metric to the exact exterior Schwarzschild solution, \(g_{tt} = -(1 - 2M/r)\) and \(g_{rr} = (1 - 2M/r)^{-1}\), where \(M\) is the numerically computed ADM mass.

\par
A naive piecewise hard-matching at the boundary \(r_1\) would introduce discontinuities in the first and second derivatives of the metric. Such abrupt jumps force the ray-tracing ODE solver to infinitely reduce its step size, ultimately leading to numerical collapse. To resolve this, we employ a linear blending scheme in the transition region \(r \in [0.9r_1, r_1]\). Specifically, the metric functions in this region are constructed as:
\begin{equation}
    g(r) = g_{\text{Pad\'e}}(r) + \left[ g_{\text{Sch}}(r) - g_{\text{Pad\'e}}(r) \right] \frac{r - 0.9r_1}{0.1r_1}.
\end{equation}
The linear blend itself only provides $C^0$ continuity at its endpoints. It acts merely as a sampling device to bridge the interior and exterior regions. For \(r > r_1\), we strictly enforce the Schwarzschild metric and extend the sampling domain up to \(r=250M\) to safely cover the observer's location. Finally, a globally \(C^2\)-continuous cubic spline interpolation is constructed through the interior Pad\'e samples, the blended buffer, and the exact Schwarzschild exterior. It is this global cubic spline that provides the \(C^2\) smoothness required by the geodesic integrator, completely eliminating numerical stiffness during the ray-tracing procedure.

\section{Ray-Tracing Method and Thin Disk Model}
\label{sec:method}

\par
To simulate the shadow features of the boson star in a realistic astrophysical environment, we employ a backward ray-tracing algorithm combined with a geometrically thin, optically thin accretion disk model \cite{Luminet1979}.

\subsection{Photon Geodesics and Effective Potential}

\par
In a static, spherically symmetric spacetime with metric $g_{\mu\nu}$, photons travel along null geodesics \cite{Chandrasekhar1983, Gralla2020, Perlick2018}. Owing to the time-translational and spherical symmetries, two conserved quantities arise for photon motion: the energy $E = -p_t$ and the angular momentum $L = p_\phi$. The Hamiltonian formalism $H = \frac{1}{2} g^{\mu\nu} p_\mu p_\nu = 0$ yields the equations of motion for the photon trajectory:
\begin{equation}
    \frac{dx^\mu}{d\lambda} = g^{\mu\nu} p_\nu, \quad \frac{dp_\mu}{d\lambda} = -\frac{1}{2} \partial_\mu g^{\alpha\beta} p_\alpha p_\beta,
\end{equation}
where $\lambda$ is the affine parameter. For the metric ansatz given in Eq.~\eqref{eq:metric_ansatz}, the radial motion of a photon can be expressed in terms of the effective potential $V_{\text{eff}}(r)$:
\begin{equation}
    \left( \frac{dr}{d\lambda} \right)^2 = E^2 - V_{\text{eff}}(r), \quad V_{\text{eff}}(r) = \frac{L^2}{r^2} A(r).
\end{equation}
The photon sphere, which determines the boundary of the shadow, corresponds to the maximum of the effective potential. Setting $dV_{\text{eff}}/dr = 0$ yields the photon sphere radius $r_{\text{ps}}$, which satisfies
\begin{equation}
    r_{\text{ps}} A'(r_{\text{ps}}) - 2A(r_{\text{ps}}) = 0.
\end{equation}
We note that for all the configurations presented in this work, the effective potential $V_{\text{eff}}$ possesses a maximum. The corresponding critical impact parameter $b_c = L/E$ is given by
\begin{equation}
    b_c = \frac{r_{\text{ps}}}{\sqrt{A(r_{\text{ps}})}}.
\end{equation}
For a distant observer at $r_{\text{obs}}$, the apparent shadow radius is given by $\theta_{\text{sh}} \approx b_c / r_{\text{obs}}$.
In our numerical implementation, we place a distant observer at $r_{\text{obs}} = 200M$ in the asymptotically flat region. We construct a local zero-angular-momentum observer (ZAMO) tetrad at the observer's position \cite{Grould2016} to set the initial photon four-momenta. Rays are then traced backward from the ZAMO frame toward the boson star using the \texttt{NDSolve} routine. When a ray intersects the equatorial plane ($\theta = \pi/2$), its position and momentum are recorded to compute the observed radiation.

\subsection{Thin Disk Emission and Redshift Factor}
\par
We model the accretion disk as lying in the equatorial plane with negligible thickness and optical depth. To facilitate a fair morphological comparison with classical Schwarzschild black hole images, we set the peak radiation at the inner edge to $r = 6M$, approximately corresponding to the innermost stable circular orbit (ISCO) of a Schwarzschild black hole \cite{Shakura1973, Novikov1973}. We note that this is a benchmark assumption to isolate the strong-field lensing effects, rather than a claim that the boson star necessarily hosts an ISCO exactly at $6M$. The intrinsic emission profile of the disk is modeled after the phenomenological family introduced by Gralla, Lupsasca, and Marrone \cite{Gralla2020b}, taking the following Gaussian-like form:
\begin{equation}
    I_{\text{emit}}(r) \propto \frac{\exp\left[-\frac{1}{2}\text{arcsinh}^2\left(\frac{r - 6M}{M}\right)\right]}{\sqrt{(r - 6M)^2 + M^2}}.
\end{equation}
As photons propagate from the emitting disk to the observer, their frequencies are shifted by both gravitational redshift and the kinematic Doppler effect induced by the Keplerian rotation \cite{Cunningham1973}. The Keplerian angular velocity of a circular orbit is
\begin{equation}
    \Omega_K = \sqrt{\frac{-g_{tt}'(r)}{2r}}.
\end{equation}
The generally covariant redshift factor is defined as $g = -1/(u^\mu k_\mu)$, where $u^\mu$ is the four-velocity of the emitting fluid and $k_\mu$ is the photon four-momentum. For a static observer and Keplerian circular orbits in a static spherical spacetime, this simplifies to:
\begin{equation}
    g = \frac{\nu_{\text{obs}}}{\nu_{\text{emit}}} = \frac{\sqrt{-g_{tt}(r_{\text{emit}})} (1 - \Omega_K b)}{\sqrt{-g_{tt}(r_{\text{obs}})}}.
\end{equation}
According to Liouville's theorem, the observed specific intensity scales as $I_{\text{obs}} = g^3 I_{\text{emit}}$. Due to relativistic Doppler beaming, the side of the disk approaching the observer ($\Omega_K b > 0$) appears significantly brightened, whereas the receding side is dimmed.

\par
\begin{figure}[htbp]
\centering
\includegraphics[width=\columnwidth]{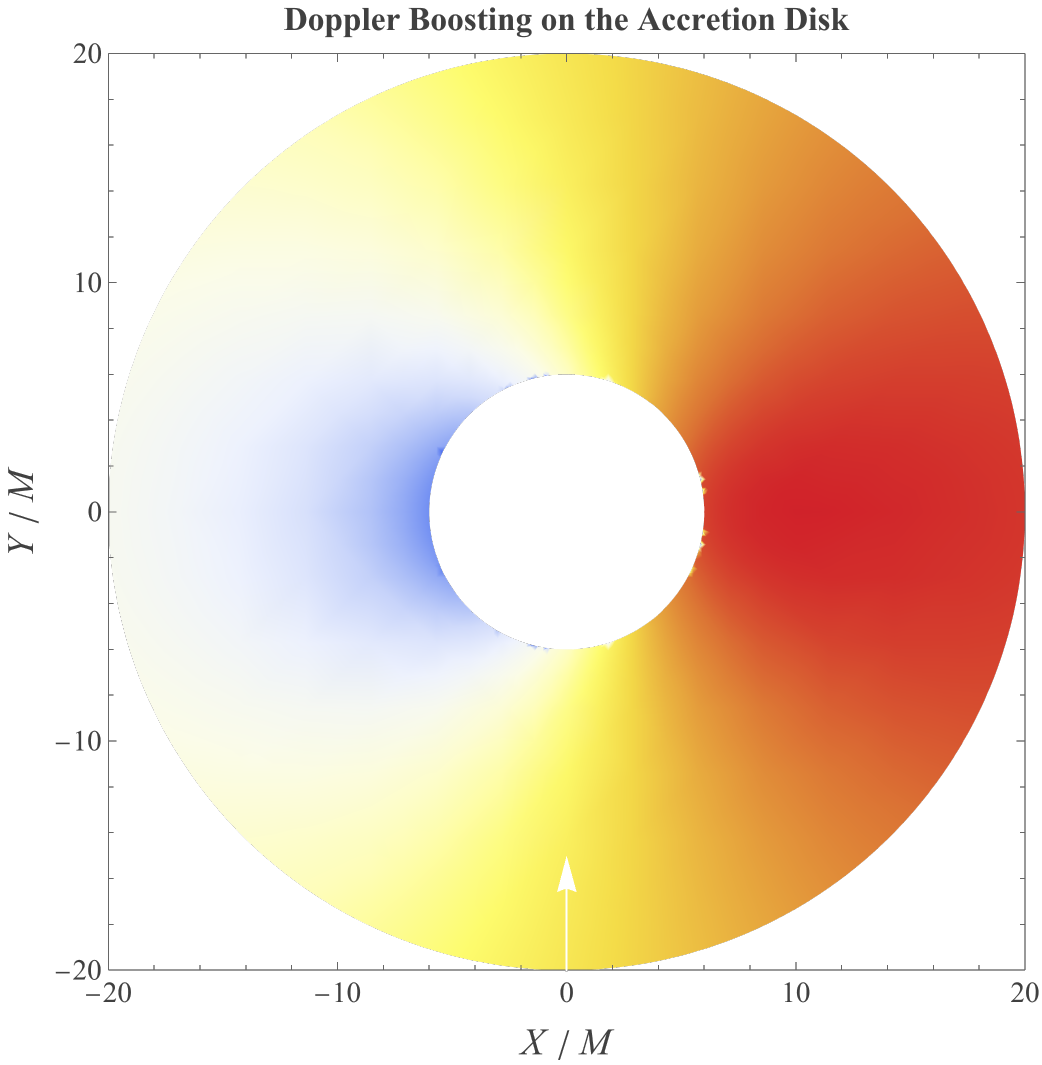}
\caption{The spatial distribution of the redshift factor $g = \nu_{\text{obs}}/\nu_{\text{emit}}$ on the equatorial plane for an observer located at an inclination angle of \(80^\circ\). The approaching side (right) exhibits strong Doppler boosting ($g>1$), appearing significantly brighter, while the receding side (left) is strongly redshifted ($g<1$). This kinematic effect is directly responsible for the asymmetric brightness observed in the shadow images.}
\label{fig:doppler_map}
\end{figure}

To visually illustrate this kinematic effect, we plot the spatial distribution of the redshift factor $g$ on the equatorial plane in Fig.~\ref{fig:doppler_map}. As clearly shown, the approaching side exhibits strong Doppler boosting ($g>1$), while the receding side is strongly redshifted ($g<1$). This pronounced asymmetric brightness distribution will serve as a dominant kinematic feature in the shadow images presented in the following sections.

\section{Numerical Results: Shadow Images}
\label{sec:results}

\par
In all ray-tracing simulations, we place the observer at a distance of $r_{\text{obs}} = 200M$ in the asymptotically flat region, with a fixed field of view (FOV) of \(5^\circ\). The accretion disk is modeled as a geometrically thin, optically thin disk in the equatorial plane, with the peak of the intrinsic emission profile located at \(r = 6M\) to mimic the radiation near the innermost stable circular orbit (ISCO) of a Schwarzschild black hole. For each pixel on the image plane, the viewing angles are mapped to Cartesian coordinates \((X, Y)\) via a stereographic projection based on the ZAMO tetrad \cite{Grould2016}. We trace a ray backward from the observer corresponding to each pixel. If the ray penetrates the equatorial plane, we accumulate the corresponding observed specific intensity \(I_{\text{obs}}\); if the ray falls into the shadow or escapes to infinity, its contribution is zero.

\par
Given the enormous dynamic range between the brightest direct emission and the faint higher-order lensed images, a linear intensity mapping would render the subtle features virtually invisible. This challenge is well known in black hole imaging \cite{Luminet1979, Johnson2020}. Therefore, to make the brightness of different images in the nine-panel grids comparable and to enhance the visibility of fine structures (such as the secondary images and higher-order photon rings), we employ a global maximum intensity normalization. We compute the global maximum intensity $I_{\max}$ across all images and apply a non-linear gamma correction to the normalized intensity \cite{Grould2016}:
\begin{equation}
    I_{\text{display}} = \left( \frac{I_{\text{obs}}}{I_{\max}} \right)^\gamma,
\end{equation}
where we set $\gamma = 0.6$. This gamma compression effectively lifts the dark regions without overexposing the bright ones, allowing the relative brightness of the shadows to be compared directly under a unified color scale. Finally, a custom colormap is applied to visualize the normalized intensity.

\par
To quantitatively describe the shadow size, we recall that for a static spherically symmetric spacetime, the apparent shadow radius $\theta_{\text{sh}}$ as seen by a distant observer is determined by the critical impact parameter $b_c$, as introduced in Sec.~\ref{sec:method}. The shadow size is highly sensitive to both the mass $M$ and the metric function \(g_{tt}\) near the photon sphere.

\subsection{Dependence on the Central Scalar Field Amplitude}
\label{sec:phi_dependence}
\par
\begin{figure*}[htbp]
\centering
\includegraphics[width=0.65\textwidth]{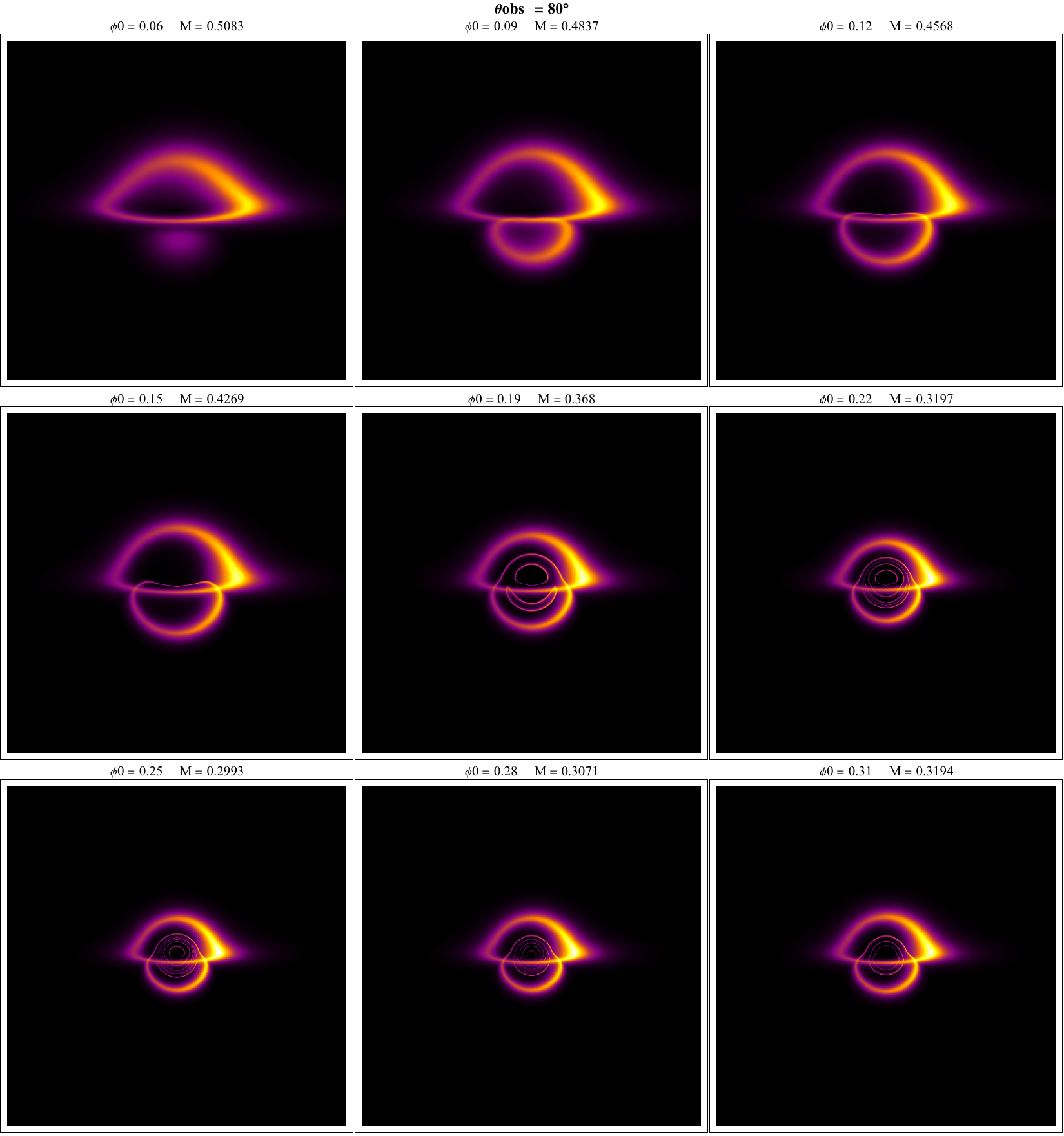}
\caption{Shadow images for nine different central scalar field amplitudes $\phi_0 \in [0.06, 0.31]$ at a fixed observation inclination of $\theta_{\text{obs}} = 80^\circ$. The shadow size does not vary monotonically with increasing $\phi_0$. The primary image of the disk crosses the shadow, and higher-order photon rings emerge as the shadow shrinks.}
\label{fig:phi_scan}
\end{figure*}
\begin{figure*}[htbp]
\centering
\includegraphics[width=0.65\textwidth]{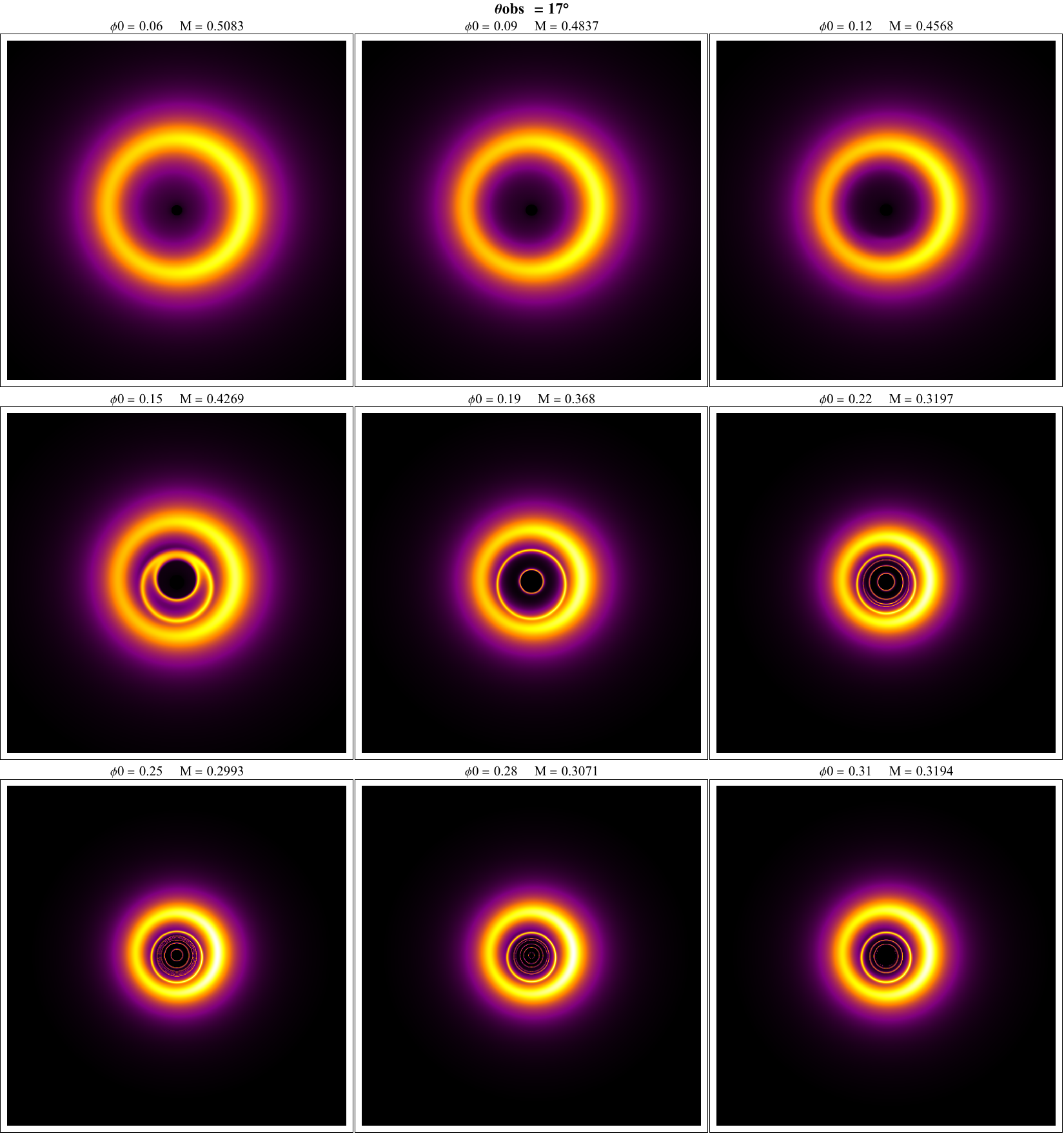}
\caption{Same as Fig.~\ref{fig:phi_scan}, but for a near face-on inclination of $\theta_{\text{obs}} = 17^\circ$. The primary and secondary images overlap into a continuous bright ring, with the higher-order photon rings prominently visible in the inner regions for compact configurations.}
\label{fig:phi_scan_17}
\end{figure*}

In this subsection we fix the self-interaction parameters at the benchmark values $\mu=1$, $\Lambda=50$, $\Gamma=625$ adopted in Sec.~\ref{sec:potential}, and vary only the central amplitude $\phi_0$. Figures~\ref{fig:phi_scan} and~\ref{fig:phi_scan_17} display the shadow images for nine different central scalar field amplitudes $\phi_0 \in [0.06, 0.31]$ at fixed observation inclinations of $\theta_{\text{obs}} = 80^\circ$ (near edge-on) and \(17^\circ\) (near face-on), respectively.

\par
Although boson stars lack an event horizon---allowing light rays with small impact parameters to theoretically traverse the stellar interior---the central region of our simulated images appears completely dark. This is driven by two effects: first, the extreme gravitational redshift near the core (as \(A_0 \to 0\)) exponentially suppresses the observed specific intensity of any traversing photons; second, the intense interior curvature strongly deflects these rays, preventing them from intersecting the luminous regions of the equatorial accretion disk.

\par
In the near edge-on view (Fig.~\ref{fig:phi_scan}), the gravitational lensing effect severely distorts the disk's primary image into a thick, asymmetric bright band that directly crosses the central shadow. The extreme relativistic Doppler beaming causes the approaching side (right) to be intensely brightened, while the receding side is highly dimmed. As $\phi_0$ increases and the shadow size changes, a series of distinct secondary and higher-order lensed images emerge below the shadow. In contrast, at the near face-on view (Fig.~\ref{fig:phi_scan_17}), the axisymmetry allows the secondary and primary images to merge seamlessly, forming a highly symmetric, continuous bright ring. For the most compact configurations, nested higher-order photon rings become distinctly visible, forming a striking geometric pattern that encodes the strong-field spacetime structure.

\par
\begin{figure}[htbp]
\centering
\includegraphics[width=\columnwidth]{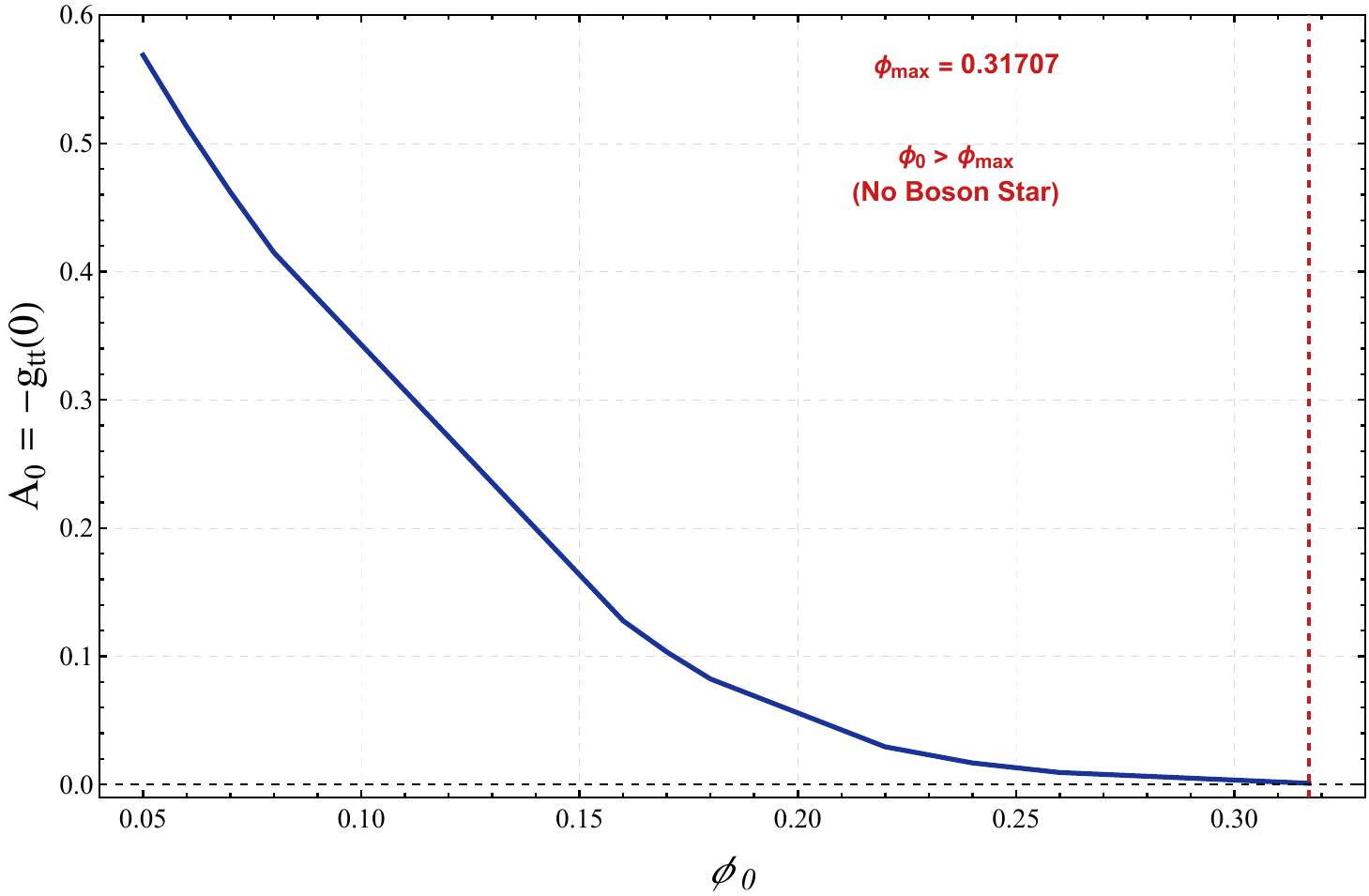}
\caption{The central metric component $A_0 = -g_{tt}(0)$ as a function of the central amplitude $\phi_0$. The red dashed vertical line marks the upper critical limit $\phi_0 \approx 0.317$. As $\phi_0$ approaches this threshold, $A_0$ drops toward zero, indicating a diverging central redshift and the loss of regular static boson-star configurations.}
\label{fig:a0_vs_phi0}
\end{figure}

Before discussing the mass turnover, it is crucial to determine the physical bounds of the central amplitude $\phi_0$. Through our high-density numerical scans, we track the evolution of the central metric component $A_0$ as a function of $\phi_0$. As shown in Fig.~\ref{fig:a0_vs_phi0}, $A_0$ decreases monotonically as the central scalar field strength increases, and regular boson star solutions abruptly terminate at a strict upper bound of $\phi_0 \approx 0.317$.

\par
As $\phi_0$ approaches this threshold, $A_0$ drops toward zero and the central gravitational redshift diverges, indicating that the repulsive $\Gamma\phi^6$ pressure can no longer support a regular static configuration. Beyond this point our shooting method ceases to converge, and no regular static boson-star solutions exist. The present equilibrium analysis does not establish the formation of an event horizon, nor that the spacetime dynamically collapses to a black hole; those questions are left for a dedicated stability and evolution study. The existence of a finite upper bound on $\phi_0$ nevertheless highlights a structural difference between boson stars and black holes: while black holes can in principle grow without a microphysical mass limit of this kind, the maximum mass and central amplitude of a regular boson star are constrained by the scalar potential. A horizon-scale object whose mass exceeded the corresponding theoretical limit of this model could not be a regular static boson star of the type considered here.

\par
\begin{figure}[htbp]
\centering
\includegraphics[width=\columnwidth]{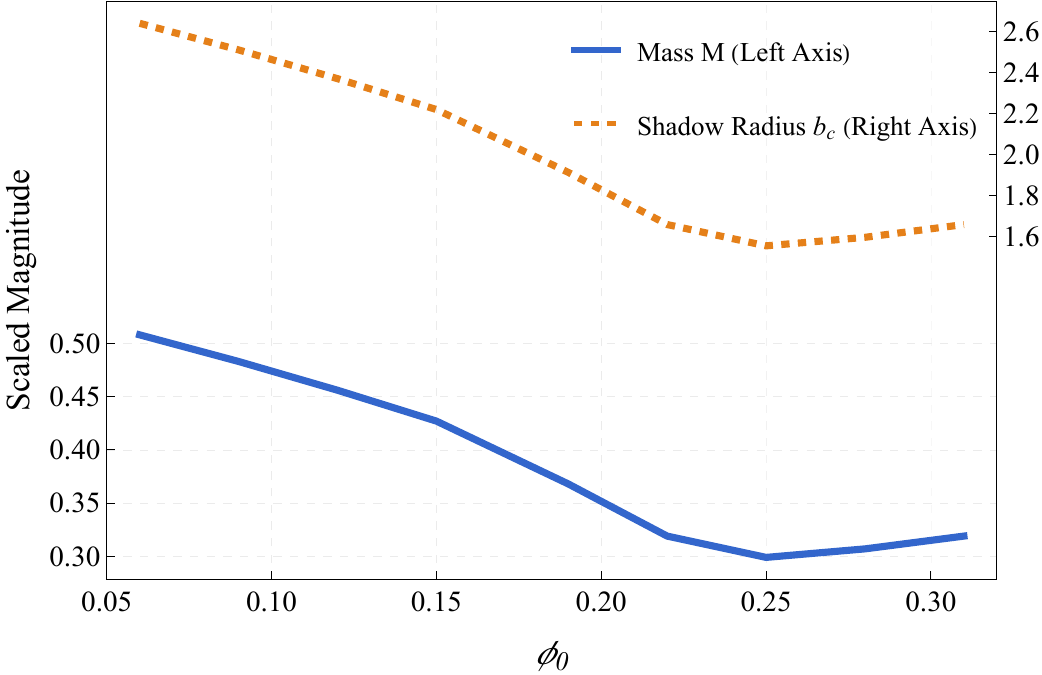}
\caption{The mass-amplitude turnover and its direct impact on the shadow radius. The blue solid line shows the ADM mass $M$ of the boson star as a function of the central field amplitude $\phi_0$ (left axis). The orange dashed line represents the corresponding shadow radius (critical impact parameter $b_c$) in the same parameter space (right axis). Both curves exhibit a synchronous non-monotonic ``turnover'' behavior, reaching a minimum at $\phi_0 \approx 0.25$.}
\label{fig:mass_turnover}
\end{figure}

Within the allowed parameter space, the shadow size does not vary monotonically with increasing $\phi_0$ (see Fig.~\ref{fig:mass_turnover}). This non-monotonic behavior is a direct physical consequence of the mass-amplitude turnover.

\par
As $\phi_0$ grows, the attractive self-interaction initially dominates, compressing the star and reducing its mass. However, once $\phi_0$ exceeds $0.25$, the repulsive sixth-order term $\Gamma\phi^6$ becomes significant. It provides extra outward pressure that halts further compression and causes the mass to slightly rebound. Because the shadow radius $b_c$ is directly tied to the ADM mass and the photon sphere geometry, it follows the same synchronous turnover. In the most compact regime ($\phi_0 \to 0.25$), a series of nested higher-order photon rings ($n \ge 1$) emerge clearly inside the central shadow. These rings correspond to light rays that orbit the boson star multiple times, providing a geometric probe of compactness. Similar higher-order photon rings have been analyzed for deformed black holes with thin accretion disks \cite{Gong2026Shadows,Guo2024KN}.

\par
It is worth noting that, according to standard stability criteria for boson stars (e.g., the Harrison-Zeldovich criterion), configurations beyond the mass-turnover point (i.e., \(\phi_0 \gtrsim 0.25\)) are typically susceptible to radial instabilities. However, our primary focus in this work is to explore the exact gravitational lensing signatures and the morphological evolution of the spacetime geometry across the entire parameter space of static solutions. A fully non-linear dynamical evolution to determine the ultimate fate of these post-turnover configurations (e.g., migration to the stable branch or collapse to a black hole) is beyond the scope of the present ray-tracing study and is left for future work.

\par
In the large-mass (dilute) regime (e.g., $\phi_0 = 0.06$), the star has a relatively large mass ($M \approx 0.508$). The shadow is large, and the disk's primary image forms a thick, highly asymmetric bright band crossing the shadow. As $\phi_0$ increases toward the mass minimum ($\phi_0 \to 0.25$), the ADM mass decreases to $M \approx 0.299$ and the shadow shrinks. For $\phi_0 \ge 0.19$, nested concentric higher-order photon rings ($n \ge 1$) become clearly visible inside the central shadow. On the post-turnover branch (e.g., $\phi_0 = 0.31$), supported by the repulsive sixth-order term, the mass $M$ rebounds slightly and the shadow expands marginally. At this stage the shadow remains highly compact, and the photon rings are packed densely together.

\subsection{Evolution with Observation Inclination}
\label{sec:angle_dependence}

\par
\begin{figure*}[htbp]
\centering
\includegraphics[width=0.65\textwidth]{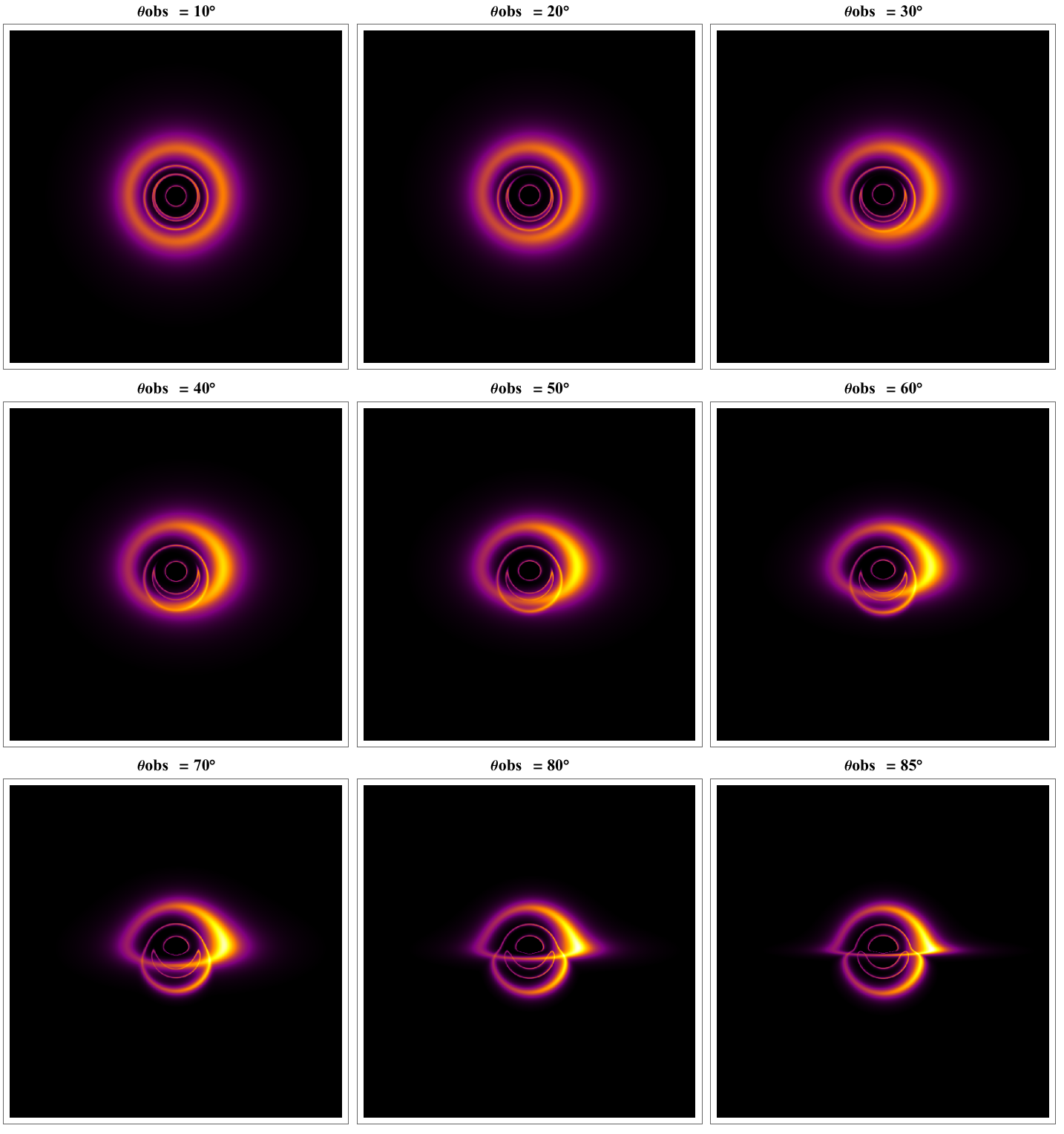}
\caption{Shadow images for nine different observation inclinations, ranging from face-on ($10^\circ$) to near edge-on ($85^\circ$). The central amplitude is fixed at the vacuum expectation value (VEV), $\phi_0 = 0.20$. The sequence clearly demonstrates how gravitational lensing distorts the disk's primary image and separates the secondary image as the viewing angle approaches the equatorial plane.}
\label{fig:angle_scan}
\end{figure*}

To systematically demonstrate how gravitational lensing alters the image with viewing angle, we fix the central amplitude at the vacuum expectation value ($\phi_0 = 0.20$) and scan nine different observation inclinations, ranging from face-on (\(10^\circ\)) to near edge-on (\(85^\circ\)), as shown in Fig.~\ref{fig:angle_scan}. The morphological evolution can be divided into three distinct regimes. For near face-on inclinations ($\theta_{\text{obs}} \in [10^\circ, 30^\circ]$), the observer looks almost directly down onto the accretion disk. Because the boson star is strictly spherically symmetric, the shadow appears as a perfect, dark circle. The intense gravitational field bends light from the far side of the disk over the top and bottom, creating a continuous, symmetric, and extremely bright photon ring that completely encircles the shadow. At intermediate inclinations ($\theta_{\text{obs}} \in [40^\circ, 60^\circ]$), the shadow remains circular, but projection effects emerge. The primary image of the disk is elongated in the viewing plane, manifesting as an asymmetric arc-shaped band. The side approaching the observer begins to show enhanced brightness due to Doppler effects.

\par
For near edge-on inclinations ($\theta_{\text{obs}} \in [70^\circ, 85^\circ]$), the primary image evolves into a highly distorted, bright band that directly crosses the shadow. The relativistic Doppler beaming effect becomes extreme, rendering the right side of the disk significantly brighter and blue-shifted compared to the left. Crucially, as the inclination increases, the secondary image (which is light from the back of the disk bent underneath the shadow) separates distinctly from the main ring and appears below the shadow. Together with the nested higher-order photon rings at compact configurations, this separated secondary image constitutes the characteristic multiple lensed rings signature of the strong gravitational field in a highly inclined system. Such multi-ring features and their morphological evolution with inclination have been extensively investigated in the thin-disk images of Kerr-Newman black holes \cite{Guo2024KN}.

\subsection{Sensitivity to the Scalar Field Potential Parameters}
\label{sec:sensitivity}

\par
\begin{figure*}[htbp]
\centering
\includegraphics[width=0.65\textwidth]{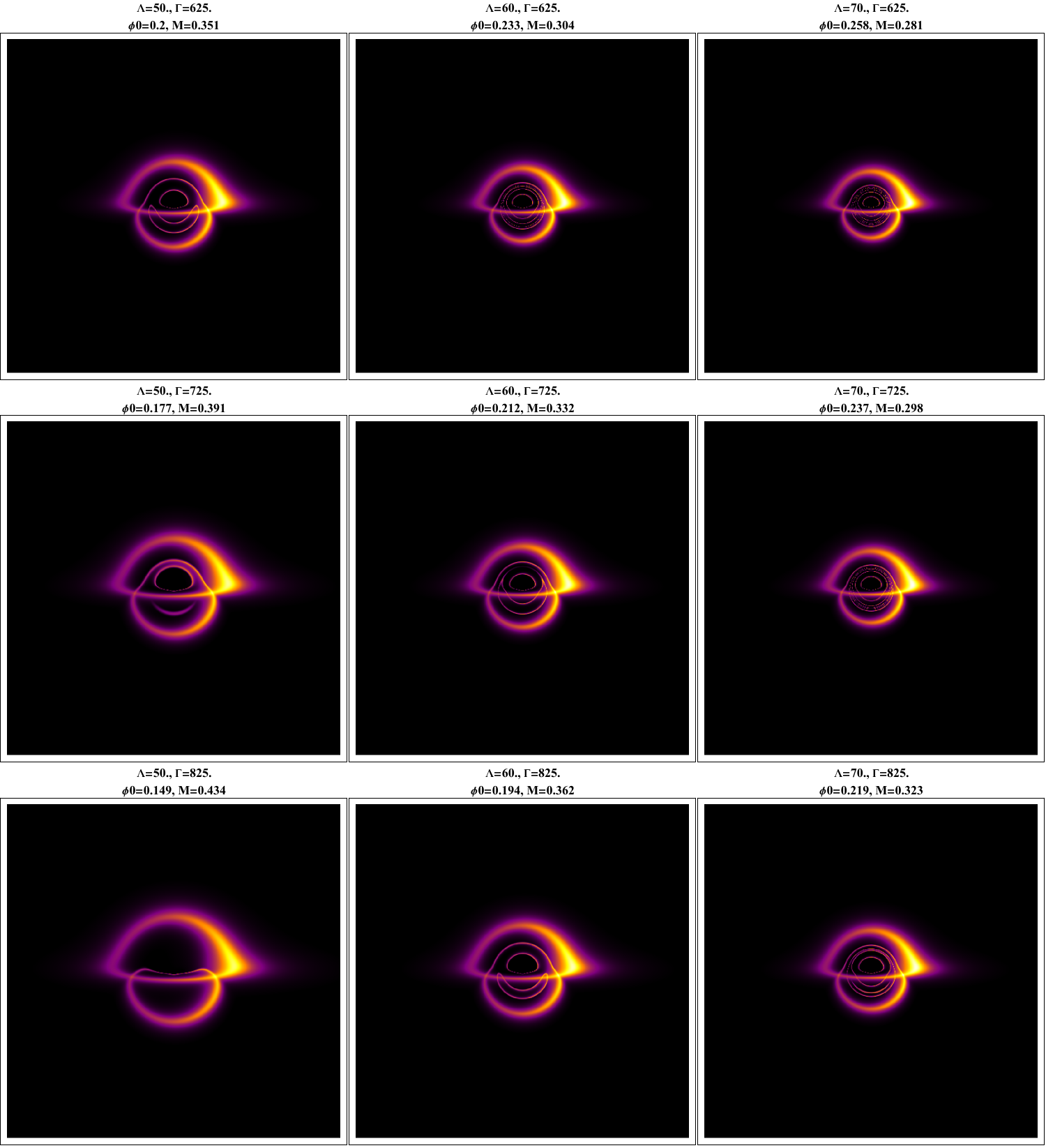}
\caption{A $3 \times 3$ grid of shadow images for varying self-interaction parameters: $\Lambda \in \{50, 60, 70\}$ (columns) and $\Gamma \in \{625, 725, 825\}$ (rows). The corresponding dynamically determined central amplitude $\phi_0$ and ADM mass $M$ are annotated in each panel. The shadow size is highly sensitive to the potential parameters, exhibiting a distinct diagonal gradient from the bottom-left (largest mass) to the top-right (smallest mass).}
\label{fig:lambda_compare}
\end{figure*}

Finally, we explore the impact of the scalar field potential parameters on the macroscopic morphology of the shadow. In Fig.~\ref{fig:lambda_compare}, we present a $3 \times 3$ grid of shadow images by varying the self-interaction parameters: $\Lambda \in \{50, 60, 70\}$ across the columns and $\Gamma \in \{625, 725, 825\}$ across the rows. The central amplitude $\phi_0$ is dynamically adjusted to the true vacuum expectation value for each parameter pair to ensure the existence of regular solutions, and the observation inclination is fixed at \(80^\circ\).

\par
Figure~\ref{fig:lambda_compare} reveals a distinct diagonal gradient in the shadow size. The shadow is largest in the bottom-left panel ($\Lambda=50, \Gamma=825$) and smallest in the top-right panel ($\Lambda=70, \Gamma=625$). This gradient is a direct physical manifestation of the competition between attractive and repulsive forces.

\par
When scanning downward along a column, an increasing repulsive parameter $\Gamma$ compresses the scalar field and pushes $\phi_0$ to lower values. As established in Fig.~\ref{fig:mass_turnover}, operating on the branch where $\phi_0 < 0.20$ yields a higher ADM mass, leading to a larger critical impact parameter and a larger shadow. Conversely, when scanning rightward along a row, an increasing attractive parameter $\Lambda$ deepens the potential well and pushes $\phi_0$ to higher values. Near the turnover minimum, this higher $\phi_0$ actually reduces the ADM mass, shrinking the shadow and making the higher-order photon rings more prominent.

\par
These morphological differences highlight the dual potential of future high-precision VLBI observations (e.g., the ngEHT and space-based VLBI arrays). They provide a powerful tool to test modified gravity theories and constrain alternative compact objects through precise shadow templates \cite{Kuang2021,Kuang2022}. Moreover, comparing our theoretical templates with precise shadow measurements and mass estimates \cite{EHT2019VI} allows us to directly probe and constrain the microphysical self-interaction parameters of dark matter scalar fields \cite{Gomez2024}, observationally distinguishing between different dark matter models in the strong-field regime.

\FloatBarrier

\section{Conclusions}
\label{sec:conclusions}
\par
In this paper, we have investigated the non-monotonic shadow sizes and multiple lensed rings of self-interacting boson stars illuminated by a thin accretion disk. To overcome the severe numerical challenges from the slowly decaying scalar tail, we developed a robust hybrid metric reconstruction framework. By connecting an interior 6-parameter Pad\'e approximant to the exact exterior Schwarzschild metric via $C^2$-continuous cubic spline interpolation, we ensure high-precision ray-tracing and completely eliminate the solver collapse that plagues traditional approaches.

\par
Importantly, the hybrid Pad\'e-spline metric reconstruction framework developed here provides a robust and generalizable numerical tool. It can be readily applied to high-precision ray-tracing in other modified gravity or dark matter models featuring slowly decaying fields, where traditional piecewise matching inevitably causes solver breakdown.

\par
Using this rigorous framework, our results reveal fundamental physical properties of these dark matter candidates. First, for the fixed self-interaction parameters adopted in this work, we establish a strict upper bound ($\phi_0 \lesssim 0.317$) for regular boson star solutions. Beyond this limit, no regular static solutions exist, as the central metric component $A_0$ approaches zero, the central redshift diverges, and the repulsive $\Gamma\phi^6$ pressure is insufficient to balance self-gravity. Whether such configurations dynamically collapse to a black hole is left for a dedicated stability analysis. More importantly, we identify a mass-amplitude turnover driven by the intense competition between attractive and repulsive self-interactions. This microphysical turnover directly dictates a pronounced non-monotonic behavior in the macroscopic shadow size, breaking the conventional monotonic paradigm. Near the turnover minimum, multiple lensed rings---including nested higher-order photon rings and distinctly separated secondary images---become highly visible, serving as a direct geometric probe of the star's compactness. Furthermore, varying the potential parameters ($\Lambda, \Gamma$) produces a sensitive diagonal gradient in the shadow size, providing a unique multidimensional space for observational constraints.

\par
These features carry important observational implications. The non-monotonic shadow variations and multiple lensed rings provide concrete, testable theoretical templates for future high-precision VLBI observations. It should be acknowledged that extracting these infinitely thin higher-order photon rings from actual observational data remains highly challenging due to the finite interferometric resolution and blurring effects of current arrays. Nevertheless, comparing these pristine theoretical templates with future data from next-generation arrays, including ngEHT and space-based VLBI, will allow us not only to identify horizonless compact objects but also to directly constrain the microphysics of dark matter scalar fields---features that cannot be obtained from black hole shadows alone. While our current model assumes a static, spherically symmetric configuration and a simplified thin-disk profile, extending this hybrid ray-tracing framework to rotating boson stars, polarized radiative transfer, and GRMHD accretion flows will be essential for fully unlocking the potential of horizon-scale imaging in the era of precision dark matter astronomy.

\begin{acknowledgments}
This work was supported by the Bagui Scholars Programme (W.X.-G., grant No. GXR-6BG2424001).
\end{acknowledgments}

\bibliographystyle{apsrev4-2}
\bibliography{refs}

\end{document}